\documentclass[a4paper,12pt]{article}
\pdfoutput=1 

\usepackage{jcappub} 

\usepackage{bm}
\usepackage{indentfirst}
\usepackage{amsmath}
\usepackage{graphicx}
\usepackage{float}
\usepackage{amssymb}
\usepackage{subfigure}
\usepackage{amssymb}
\usepackage{psfrag}
\usepackage{epstopdf}
\usepackage{diagbox}

\title{Primordial black holes and scalar-induced gravitational waves from  the generalized Brans-Dicke theory}

\author{Zhu Yi$^{1}$}

\affiliation{$^1$Advanced Institute of Natural Sciences, Beijing Normal University, Zhuhai 519087, China}

\emailAdd{yz@bnu.edu.cn}

\abstract{The power spectrum of  the scalar-tensor inflation with a quadratic form 
	Ricci scalar  coupling function 
	$\Omega(\phi)=1- 2\phi/\phi_c+(1+\delta^2)(\phi/\phi_c)^2$  can be enhanced enough to produce primordial black holes  and generate scalar-induced gravitational waves. The masses of primordial black holes and the frequencies of scalar-induced gravitational waves are controlled by the parameter $\phi_c$, and their amplitudes  are  determined by the parameter $\delta$.  Primordial black holes with stellar masses, planetary masses, and masses around $10^{-12} M_\odot$ are produced and their abundances are obtained from  the peak theory.  
	The frequencies of  the corresponding scalar-induced gravitational waves are around $10^{-9}$ Hz, $10^{-6}$ Hz, and $10^{-3}$ Hz, respectively.  
	The primordial black holes with masses around $10^{-12} M_\odot$ can account for almost all of the dark matter, and the scalar-induced gravitational waves  with  frequencies around $10^{-9}$ Hz can explain the NANOGrav 12.5yrs signal.}

\begin{document}
	\maketitle
	\flushbottom

\section{Introduction}
Primordial black holes (PBHs) can be formed from the gravitational collapse of overdense regions with their density contrasts exceeding the threshold value at the horizon reentry during radiation domination \cite{Carr:1974nx,Hawking:1971ei}. 
PBHs with stellar masses may be   the black holes in the
gravitational waves (GWs) events   detected by the Laser  Interferometer Gravitational Wave Observatory (LIGO)  Scientific  Collaboration and the Virgo Collaboration
\cite{Bird:2016dcv,Sasaki:2016jop, Abbott:2016blz,Abbott:2016nmj, Abbott:2017vtc, Abbott:2017oio,TheLIGOScientific:2017qsa, Abbott:2017gyy,LIGOScientific:2018mvr,Abbott:2020uma,LIGOScientific:2020stg,  Abbott:2020khf,Abbott:2020tfl,LIGOScientific:2020ibl}.  
PBHs with planetary masses  can explain  the ultrashort-timescale microlensing events in the OGLE data \cite{Niikura:2019kqi}, 
and can act as the Planet 9 which is a hypothetical astrophysical object in the outer solar system used to explain  the anomalous orbits of trans-Neptunian objects \cite{Scholtz:2019csj}. 
PBHs are also proposed to account for dark matter (DM) \cite{Ivanov:1994pa,Frampton:2010sw,Belotsky:2014kca,Khlopov:2004sc, Carr:2016drx,Inomata:2017okj,Garcia-Bellido:2017fdg,Kovetz:2017rvv,Carr:2020xqk},
and  those  with masses  around $10^{-17}-10^{-15} M_{\odot}$ and $10^{-14}-10^{-12}M_{\odot}$ can make up almost all of DM
for there are no  observational constraints on  the abundances of PBHs at these mass windows.  

The overdense regions, collapsing to  PBHs by  gravitational force,  originate  from  the primordial curvature perturbations generated during Inflation.  
From the threshold value of the  density contrasts for PBHs formation, the amplitude of the power spectrum  of the primordial curvature perturbations is constrained to  $A_\zeta\sim \mathcal{O}(0.01)$, which is seven orders of magnitude larger  \cite{Gong:2017qlj,Lu:2019sti,Sato-Polito:2019hws} than the large scale  constraints $A_\zeta=2.1\times 10^{-9}$ \cite{Akrami:2018odb} from the observation of  cosmic microwave background (CMB) anisotropy measurements.  
Therefore, the allowed way to  produce enough PBHs DM is by   enhancing the power spectrum by  about seven orders of magnitude at small scales.

The traditional slow-roll inflation model is  hard to enhance the power spectrum  at small scales
while keeping the model consistent with the large scale  constraints.
To solve this difficult, we need consider the ultra-slow-roll inflation model  that transiently  satisfies the  condition $\ddot{\phi}+3 H\dot{\phi}\approx 0$ \cite{Martin:2012pe,Motohashi:2014ppa,Yi:2017mxs}.  For the canonical inflation models with a single field, a simple way to realize  the ultra-slow-roll inflation is by introducing  an inflection point in  the potential \cite{Garcia-Bellido:2017mdw,Germani:2017bcs,Motohashi:2017kbs,Ezquiaga:2017fvi,Gong:2017qlj,Ballesteros:2018wlw,Dalianis:2018frf,Bezrukov:2017dyv,Kannike:2017bxn}. 
However,  it is not easy to achieve the big enhancement on the power spectrum while keeping  the total number of e-folds
around $50-60$  \cite{Sasaki:2018dmp,Passaglia:2018ixg}. Noncanonical kinetic terms inflation \cite{Lin:2020goi,Lin:2021vwc,Gao:2020tsa,Gao:2021vxb,Yi:2020kmq,Yi:2020cut,Yi:2021lxc,Zhang:2020uek,Pi:2017gih} or other kinds of noncanonical  inflation models \cite{Kamenshchik:2018sig,Fu:2019ttf,Fu:2019vqc,Dalianis:2019vit,Gundhi:2020zvb,Cheong:2019vzl,Zhang:2021rqs,Kawai:2021edk,Cai:2021wzd,Chen:2021nio,Zheng:2021vda,Karam:2022nym,Ashoorioon:2018uey,Ashoorioon:2019xqc} were then considered. 
For example, with the coupling function $f(\phi)$ and potential satisfying $(V_\phi+V^2 f_\phi/6 )|_{\phi=\phi_c}\approx0$, the Gauss-Bonnet inflation model has a  transient ultra-slow-roll process at the critical point $\phi_c$ \cite{Zhang:2021rqs,Kawai:2021edk} and succeeds  in enhancing the power spectrum and produce PBHs.  The noncanonical kinetic term  inflation
model with  coupling function $G(\phi)=h/[1+\left(|\phi-\phi_c|/{w}\right)^q]$ can realize a large  enhancement on the power spectrum  and produce PBHs if the parameter $h$ is large enough \cite{Lin:2020goi,Yi:2020cut}.  
Besides the noncanonical single field inflation models, the multi-filed inflationary models 
are another important way to enhance the power spectrum \cite{Garcia-Bellido:1996mdl,Clesse:2015wea,Palma:2020ejf},  especially those 
with tachyonic instabilities \cite{Braglia:2020eai,Fumagalli:2020nvq,Cheong:2022gfc}.
In this paper, we focus on the scalar-tensor inflation  and find that with the  Ricci scalar  coupling function being a quadratic form  $\Omega(\phi)=1- 2\phi/\phi_c+(1+\delta^2)(\phi/\phi_c)^2$, 
the ultra-slow-roll condition can be satisfied  transiently at the critical point $\phi_c$, and the power spectrum can be enhanced enough to produce PBHs.  The masses  and abundances of the PBHs can be adjusted by the parameters $\phi_c$ and $\delta$, respectively. 
It was pointed out recently that a single field inflation model,  producing an appreciable amount of PBHs, is in danger of excessive one-loop corrections to the   CMB scale  \cite{Kristiano:2022maq}. Hence, our model is potentially in danger of the one-loop effect, but it is beyond the scope of the present paper and left for future study.

With  the formation of PBHs, the large scalar perturbations at small scales induce secondary gravitational waves  after the horizon reentry during  the radiation dominated epoch \cite{Matarrese:1997ay,Mollerach:2003nq,Ananda:2006af,Baumann:2007zm,Garcia-Bellido:2017aan,Saito:2008jc,Saito:2009jt,Bugaev:2009zh,Bugaev:2010bb,Alabidi:2012ex,Orlofsky:2016vbd,Nakama:2016gzw,Inomata:2016rbd,Cheng:2018yyr,Cai:2018dig,Bartolo:2018rku,Bartolo:2018evs,Kohri:2018awv,Espinosa:2018eve,Cai:2019amo,Cai:2019elf,Cai:2019bmk,Cai:2020fnq,Domenech:2019quo,Domenech:2020kqm,Fumagalli:2020adf,Ashoorioon:2020hln,Ashoorioon:2022raz,Pi:2020otn,Yuan:2019fwv,Yuan:2019wwo,Yuan:2019udt,Papanikolaou:2020qtd,Papanikolaou:2021uhe,Papanikolaou:2022hkg,Domenech:2021ztg,Atal:2021jyo,Balaji:2022dbi,Feng:2023veu,Zhang:2022xmm}. 
These scalar-induced gravitational waves (SIGWs) have wide frequency distribution and can be detected by  pulsar timing arrays (PTA) \cite{Ferdman:2010xq,Hobbs:2009yy,McLaughlin:2013ira,Hobbs:2013aka,Moore:2014lga}
and the space-based GW detectors such as Laser Interferometer Space Antenna (LISA) \cite{Danzmann:1997hm,Audley:2017drz}, Taiji \cite{Hu:2017mde},  TianQin  \cite{Luo:2015ght} and   Deci-hertz Interferometer Gravitational Wave Observatory (DECIGO) \cite{Kawamura:2011zz} in the future.  
For example, the stochastic process  with a common amplitude and a common spectral slope across  pulsars detected by 
the North American Nanohertz Observatory for Gravitational Wave (NANOGrav)  Collaboration \cite{NANOGrav:2020bcs} and other pulsar
timing arrays  \cite{Goncharov:2021oub,Antoniadis:2022pcn}  recently 
may be the SIGWs with nHz frequencies\cite{DeLuca:2020agl,Inomata:2020xad, Vaskonen:2020lbd, Domenech:2020ers, Yi:2021lxc, Yi:2022ymw}. 

The paper is organized as follows. In Sec. II,  we show the enhancement mechanism on the power spectrum of  the scalar-tensor inflation in detail.
We discuss the production of PBH DM and the generation of SIGWs from this mechanism in Sec. III.
We conclude the paper in Sec. V.

\section{The model}
The action for the scalar-tensor theory in the Jordan frame is 
\begin{equation}\label{actjd}
	S=\int \sqrt{-g}dx^4\left[ \frac{\Omega(\phi)}{2}R
	-\frac{\omega(\phi)}{2}(\partial\phi)^2-V(\phi)\right],
\end{equation}
where $\Omega(\phi)$  and $\omega(\phi)$ are  the coupling functions,  and  $V(\phi)$ is the potential, $(\partial\phi)^2=g^{\mu\nu}\nabla_\mu\phi\nabla_\nu\phi$. The reduced Planck mass is  $M_{\rm pl}=1/\sqrt{8\pi G}$, and the units  are $c=\hbar=M_{\rm pl}=1$.
For the homogeneous and isotropic background,   the Friedmann equation and the equation of motion for the scalar field are
\begin{equation}
	\label{bgeq1}
	H^2=\frac{1}{3\Omega}\left(\frac{1}{2}\omega  \dot{\phi}^2+V\right)-H \frac{\dot{\Omega}}{\Omega},
\end{equation}
\begin{equation}
\label{bgeq2}
\begin{aligned}
&\left[\omega+\frac{3}{2\Omega}\left( \frac{d\Omega}{d\phi}\right)^2\right](\ddot{\phi}+3H\dot{\phi})+\frac{dV}{d\phi}-\frac{2V}{\Omega}\frac{ d  \Omega}{d\phi}\\
&\quad +\left(\frac{3}{2\Omega}\frac{d\Omega}{d\phi}\frac{d^2\Omega}{d\phi^2}+\frac{1}{2}\frac{d\omega}{d\phi}+\frac{1}{\Omega}\frac{d\Omega}{d\phi}\frac{w}{2}\right)\dot{\phi}^2=0,
\end{aligned}
\end{equation}
where a ``dot'' denotes the derivative with respect to cosmic time $t$.
Under the slow-roll conditions \cite{Torres:1996fr}
\begin{equation}
\frac{1}{2}\omega(\phi)	\dot{\phi}^2\ll V(\phi) ,\quad |\dot{g}|\ll H g,
\end{equation}
where the function $g$ denotes an arbitrary function, such as $\Omega(\phi)$ and $\omega(\phi)$, the background equations \eqref{bgeq1} and \eqref{bgeq2} become
\begin{gather}
	\label{slbgeq1}
	H^2\approx \frac{V(\phi)}{3\Omega(\phi)},\\
	\label{slbgeq2}
	(\ddot{\phi}+3H\dot{\phi})+\Gamma(\phi)\frac{dU(\phi)}{d\phi}\approx 0,
\end{gather}
with the effective potential $U(\phi)=V(\phi)/\Omega(\phi)^2$  and 
\begin{equation}\label{defA}
\Gamma(\phi)=\frac{\Omega(\phi)^2}{\omega(\phi)+\frac{3}{2\Omega}\left(\frac{d\Omega}{d\phi}\right)^2}.
\end{equation}
The condition of  forming  PBHs  requires the amplitude of  the power spectrum of the primordial curvature to reach around $A_\zeta\sim \mathcal{O}(0.01)$, while the  constraints on power spectrum at  large scales from the observation of CMB  anisotropy measurements is $A_\zeta=2.1\times 10^{-9}$ \cite{Planck:2018jri}. Therefore, to produce PBHs, the  power spectrum should be enhanced by about seven orders of magnitude at small scales, and this  is  hard to realize in the slow-roll inflation model.   For the ultra-slow-roll inflation with  condition \cite{Martin:2012pe}
\begin{equation}\label{usleq}
	\ddot{\phi}+3H\dot{\phi} \approx 0,
\end{equation}
the power spectrum  can be  enhanced enough   to produce PBHs. From   equation  \eqref{slbgeq2},  to obtain the ultra-slow-roll condition  \eqref{usleq},  we need
\begin{equation}\label{uslcon}
	\Gamma(\phi)\frac{dU(\phi)}{d\phi} \ll 1.
\end{equation}
If the effective potential $U(\phi)$ has a  near inflection point, $dU(\phi)/d\phi|_{\phi=\phi_c}\approx0$, the condition \eqref{uslcon} can be satisfied easily \cite{Garcia-Bellido:2017mdw,Gong:2017qlj}.   In addition to the method of  near inflection point, the other way to obtain   condition \eqref{uslcon} is by making  the coupling functions satisfy
\begin{equation}\label{uslcon2}
\Gamma(\phi)=\frac{\Omega(\phi)^2}{\omega(\phi)+\frac{3}{2\Omega}\left(\frac{d\Omega}{d\phi}\right)^2}\ll1,
\end{equation}
which requires $\omega(\phi)\gg1$ or $\Omega(\phi)\ll1$ at the ultra-slow-roll point. The situation $\omega(\phi)\gg 1$ has been researched in   papers \cite{Lin:2020goi,Yi:2020cut}  with the form
\begin{equation}\label{cou:om}
  \omega(\phi)=1+\frac{h}{1+\left(|\phi-\phi_c|/{w}\right)^q},
\end{equation}
and $  h\gg 1$. At the point $\phi_c$, the coupling function satisfies $\omega(\phi_c)=1+  h\gg1$, and the ultra-slow-roll condition \eqref{uslcon2} is satisfied.

In this paper, we consider the  other case, $\Omega(\phi)\ll 1$. Inspired by the Higgs inflation  \cite{Bezrukov:2007ep} with the coupling function  $\Omega(\phi)=1 +\xi  \phi^2$,   we consider the   second-order polynomial 
\begin{equation}\label{defOm}
	\Omega(\phi)=1- \frac{2\phi}{\phi_c} +(1+\delta^2)\left(\frac{\phi}{\phi_c}\right)^2,
\end{equation}
where the term $\delta^2\ll1$  is used to keep the coupling function $\Omega(\phi)$ from exact zero. 
At the critical point, the coupling function becomes $\Omega(\phi_c)=\delta^2\ll1$,  the condition \eqref{uslcon2}  is satisfied, the inflaton evolves into a transitory ultra-slow-roll phase where the power spectrum of the curvature perturbations is enhanced.  The critical point $\phi_c$   controls the position of the peak in the power spectrum and $\delta$ determines the amplitude of the peak.  
The  choice \eqref{defOm} may be the simplest form containing a critical point  satisfying condition  $\Omega(\phi_c)\ll1$,  and this choice   could be regarded as a phenomenological step, and finding the corresponding UV theory will be the next step.  

To obtain the condition \eqref{uslcon2}, even if the  numerator of relation \eqref{uslcon2} is chosen as equation \eqref{defOm} and satisfies $\Omega(\phi_c)\ll1$, 
the  denominator  should satisfy
\begin{equation}\label{defA2:1}
\omega(\phi_c)+\frac{3}{2\Omega(\phi_c)}\left(\frac{d\Omega(\phi_c)}{d\phi}\right)^2 \gtrsim\Omega(\phi_c)^p,
\end{equation}
at the critical point with $p<2$.  In order to satisfy condition \eqref{defA2:1}, we choose the kinetic coupling function as 
\begin{equation}\label{defA2}
\omega(\phi)+\frac{3}{2\Omega(\phi)}\left(\frac{d\Omega(\phi)}{d\phi}\right)^2=a \Omega(\phi)^{b} + c \Omega(\phi),
\end{equation}
with $b<1$, and the form of $\omega(\phi)$ can be obtained completely by that of $\Omega(\phi)$. 
The first term $a \Omega(\phi)^{b}$ is used to adjust the shape of the peak, and the second term $c \Omega(\phi)$ is used to keep $w(\phi)=1$ at the lower energy scales $\phi\ll1$, which requires
\begin{equation}
	c=1-a+\frac{6}{\phi_c^2}.
\end{equation}
The abilities to produce different  peaks in the primordial power spectrum as displayed in  figure \ref{fig:pr} 
and recover to the canonical  situation at the lower energy are the main reason to choose form \eqref{defA2}, although it is not conventional.

The potential is \cite{Lin:2021vwc}
\begin{equation}\label{potential}
	V(\phi)=\frac{\lambda \phi^4}{4} \left(\frac{\Omega(\phi)}{1+\xi\phi^2}\right)^2,
\end{equation}
with $\xi=10$. At the lower energy scales $\phi\ll1$, the potential reduces to the Higgs potential with the form $\lambda\phi^4/4$.

In the other hand, taking   the conformal transformation,
\begin{equation}
\tilde{g}_{\mu\nu}	=\Omega(\phi) g_{\mu\nu},
\end{equation}
and changing the Jordan frame to the Einstein frame,  action \eqref{actjd}  becomes
\begin{equation}
S=\int d^4x \sqrt{-\tilde{g}}\left[\frac{\tilde{R}}{2}-\frac{1}{2}k(\phi)(\tilde{\partial}\phi)^2-\frac{V(\phi)}{\Omega(\phi)^2}\right],
\end{equation}
with  $(\tilde{\partial}\phi)^2=\tilde{g}^{\mu\nu}\nabla_\mu\phi\nabla_\nu\phi$ and
\begin{equation}
k(\phi)=\frac{3}{2}\left(\frac{d\Omega/ d\phi}{\Omega}\right)^2+\frac{\omega(\phi)}{\Omega(\phi)}.
\end{equation}
Combining equation \eqref{defOm} and \eqref{defA2}, the coupling function of the kinetic term becomes
\begin{equation}\label{kphi}
	k(\phi)=1+\left(-a+\frac{6}{\phi_c^2}\right)+ \frac{a}{\Omega^{1-b}}=1+G(\phi)
\end{equation}
with 
\begin{equation}
G(\phi)=\left(\frac{6}{\phi_c^2}-a\right)+\frac{a (\phi_c/\delta)^{2-2b}}{\left[\left(\frac{\phi-\phi_c}{\delta}\right)^2+\phi^2\right]^{1-b}},
\end{equation}
which is similar with  equation \eqref{cou:om}. As pointed out in Refs. \cite{Lin:2020goi,Yi:2020cut}, this kind of kinetic term can succeed in enhancing the power spectrum and producing enough PBHs.

\section{The results}
\subsection{power spectra}
The quadratic action for the curvature perturbation $\zeta$ of the scalar-tensor inflation \eqref{actjd} is 
\begin{equation}
	S^{(2)}=\frac{1}{2}\int d\eta d^3x \frac{a^2[\omega\dot{\phi}^2+3\dot{\Omega}^2/2\Omega]}{(H+\dot{\Omega}/2\Omega)^2}\left[(\zeta')^2-(\vec{\nabla}\zeta)^2\right],
\end{equation}
where $\eta=\int dt/a$ is the conformal time and  $\zeta'=d\zeta/d\eta$. The equation for the curvature perturbation in $k$-space is
\begin{equation}\label{pereq}
	\frac{d^2 u_k}{d\eta^2}+\left(k^2-\frac{1}{z}\frac{d^2 z}{d\eta^2}\right)u_k=0,
\end{equation}
with  $u_k=z \zeta_k$ and  \cite{Hwang:1996bc}
\begin{equation}
	z^2=\frac{a^2[\omega\dot{\phi}^2+3\dot{\Omega}^2/2\Omega]}{(H+\dot{\Omega}/2\Omega)^2}.
\end{equation}
The power spectrum of the curvature perturbation is 
\begin{equation}
\mathcal{P}_\zeta=\frac{k^3}{2\pi^2}\left|\zeta_k\right|^2,
\end{equation}
which can be obtain by solving the background equations \eqref{bgeq1} and \eqref{bgeq2}, and the perturbation equation \eqref{pereq}.

\begin{figure}[htbp]
	\centering
	\includegraphics[width=0.95\columnwidth]{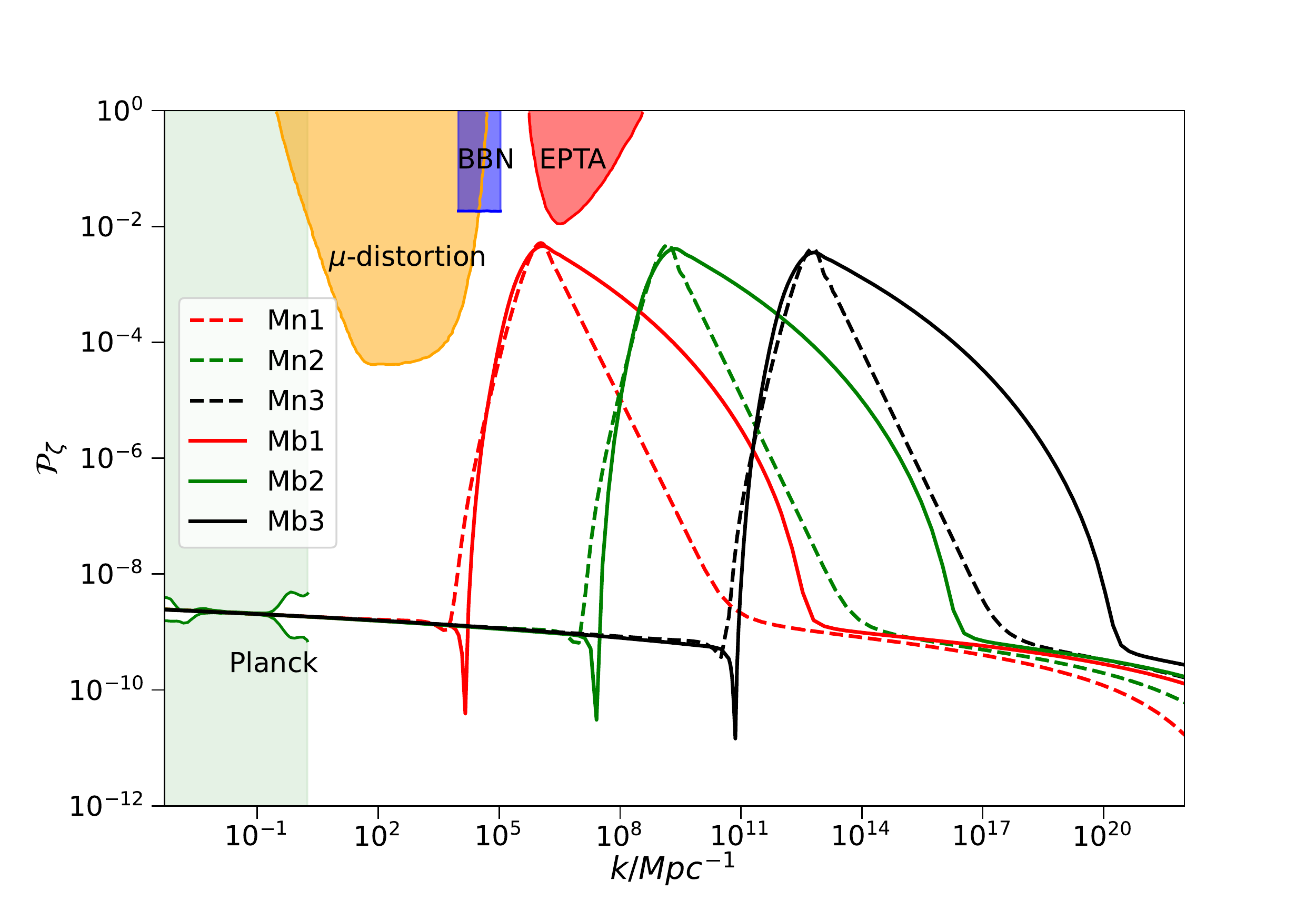}
	\caption{The power spectra of the models listed in table \ref{tab:pr}. 		
		The light green shaded region is excluded by the CMB observations \cite{Planck:2018jri}.
		The red, blue and orange regions are the
		constraints from the PTA observations \cite{Inomata:2018epa},
		the effect on the ratio between neutron and proton
		during the big bang nucleosynthesis (BBN) \cite{Inomata:2016uip}
		and $\mu$-distortion of CMB \cite{Fixsen:1996nj}, respectively.}\label{fig:pr}
\end{figure}

By choosing the values of  parameters $a$, $\delta$, $\phi_c$,  $\lambda$, and the scalar field  $\phi_*$ at the pivot scale, we can numerically obtain the power spectrum.  
For the values of the parameter  sets listed in table  \ref{tab:pr}, the numerical results of the power spectra of the curvature perturbation
 are shown in  figure \ref{fig:pr}.  The $e$-folding numbers $N$ of  these  models in table \ref{tab:pr}  are about $N \in (55, 65)$.  The scalar tilt and tensor-to-scalar ratio of these models are listed in table \ref{tab:pr} which are around
\begin{equation}
	n_s\approx 0.965, \quad r\approx 0.004,
\end{equation}
which are consistent with the observational constraints \cite{Akrami:2018odb,BICEP2:2018kqh},
\begin{gather}
	n_s=0.9649\pm 0.0042 \quad (68\%\,\text{CL}),\\
  r_{0.05}<0.06  \quad (95\%\,\text{CL}).
\end{gather}

The position of the peak of the power spectra in figure \ref{fig:pr} is controlled by the  parameter $\phi_c$ in equation \eqref{defOm}.  
The power spectra  with peak scale  around
 $k_\text{peak}\approx 10^{6} \,\text{Mpc}^{-1}$, $k_\text{peak}\approx 10^{9}\, \text{Mpc}^{-1}$, and  $k_\text{peak}\approx10^{12}\,\text{Mpc}^{-1}$ 
  are given in figure  \ref{fig:pr}, and denoted as  red  lines, green lines, and black lines, respectively.  In table \ref{tab:pr}, they are labeled as
  ``1", ``2", and ``3", respectively. The shape of the peak is determined by the index $b$ in equation \eqref{defA2}. The narrow peak denoted by the  dashed  line in  figure \ref{fig:pr}  is   from  the model with $b=1/2$ and labeled as ``Mn" in table  \ref{tab:pr},  the broad  peak denoted by the solid line in  figure \ref{fig:pr}  is   from  the model with $b=2/5$ and labeled as ``Mb" in table  \ref{tab:pr}.

\begin{table}[htbp]
	\renewcommand\tabcolsep{3pt}
		\centering
	\begin{tabular}{lllllllllll}
		\hline
		\hline
		Model \quad   &$a$& $\delta$ &$\phi_c$ &$\phi_*$&$\lambda/ 10^{-8}$ &$N$&$n_s$&$r/ 10^{-3}$&$k_{\text{peak}}/\text{Mpc}^{-1}$\\
		\hline
		Mn1 \quad   &$1.53\times 10^{-2}$~& $ 1.10\times 10^{-9}$ ~& $1.96$~&$2.13$~&$6.02$~&$54$~&$0.964$~&$4.6$~&$1.08\times 10^{6} $\\
		Mn2 \quad   &$ 1.95\times10^{-2}$~& $ 2.73\times 10^{-10}$ ~& $1.83$~&$2.13$~&$ 5.67 $~&$57$~&$0.963$~&$4.4$~&$1.52\times 10^{9} $\\
		Mn3 \quad   &$2.66\times10^{-2}$~& $ 7.93\times 10^{-11}$ ~& $1.69$~&$2.13$~&$ 5.16$~&$61$~&$0.966$	~&$4.0$~&$5.75\times 10^{12} $ \\
		\hline
		Mb1 \quad   &$1.50\times 10^{-3}$~& $ 3.18\times 10^{-11}$ ~& $2.03$~&$2.25$~&$4.77$~&$60$~&$0.965$~&$3.7$~&$1.26\times 10^{6} $\\
		Mb2 \quad   &$2.46\times10^{-3}$~& $ 1.19\times 10^{-10}$ ~& $1.82$~&$2.18$~&$ 5.05 $~&$61$~&$0.965$~&$3.9$~&$2.36\times 10^{9} $ \\
		Mb3 \quad   &$4.24\times10^{-3}$~& $ 4.67\times 10^{-11}$ ~& $1.61$~&$2.12$~&$ 5.07 $~&$65$~&$0.966$	~&$3.9$~&$6.31\times 10^{12} $\\
		\hline
		\hline
	\end{tabular}
	\caption{The chosen parameter sets and the predictions of scalar tilt $n_s$, tensor-to-scalar ratio $r$, and $e$-folds $N$.  }
	\label{tab:pr}
\end{table}

\subsection{primordial black holes}
If the amplitude of the   power spectrum  is enhanced to  $A_\zeta \sim \mathcal{O}(0.01)$ at small scales, it may form  PBHs from gravitational collapse  during the radiation domination.  The mass fraction of the Universe that collapses to form PBHs at formation is denoted by 
\begin{equation}\label{beta}
	\beta=\frac{\rho_{\text{PBH}}}{\rho_b},
\end{equation}
where $\rho_b$ is the energy density of the background and $\rho_{\text{PBH}}$ is the  energy density of the PBHs at  formation. From the peak theory, the energy density of the PBHs is  \cite{Bardeen:1985tr,Green:2004wb,Young:2014ana,Germani:2018jgr,Young:2020xmk,Gow:2020bzo},
\begin{equation}\label{rho:pbh}
	\rho_{\text{PBH}}=\int_{\nu_c}^{\infty}M_{\text{PBH}}(\nu)\mathcal{N}_{pk}(\nu)d\nu,
\end{equation}
where the number density of the PBHs is \cite{Bardeen:1985tr}
\begin{equation}\label{num:den}
	\mathcal{N}_{pk}(\nu)=\frac{1}{a^3}\frac{1}{(2\pi)^2}\left(\frac{\sigma_1}{\sqrt{3}\sigma_0}\right)^3
	\nu^3\exp\left(-\frac{\nu^2}{2}\right).
\end{equation}
The lower limit of the integral in equation \eqref{rho:pbh} is $\nu_c=\delta_c/\sigma_0$,   $\delta_c$  is the threshold for the formation of PBHs, and 
$\sigma_0$ is the variance of the smoothed density contrast. 
The moment of the smoothed density power spectrum $\sigma_1$ is defined by
\begin{equation}\label{variance1}
	\sigma^2_n=\int_{0}^{\infty}\frac{dk}{k}k^{2n} T^2(k,R_H)W^2(k,R_H)\mathcal{P}_\delta(k),
\end{equation}
where  $\mathcal{P}_\delta$ is the  power spectrum of the density contrast which is related to the 
power spectrum of  primordial curvature  perturbations $\mathcal{P}_\zeta$  by
\begin{equation}\label{rel:pp}
	\mathcal{P}_\delta(k)=\frac{4(1+w)^2}{(5+3w)^2}\left(\frac{k}{aH}\right)^4 \mathcal{P}_{\zeta}(k),
\end{equation}
with the state equation $w=1/3$ during the radiation domination. 

For the window function $W(k,R_H)$ in equation \eqref{variance1}, there are usual three choices, the  real-space top-hat window function, the Gauss window function, and the $k$-space top-hat window function  \cite{Ando:2018qdb}.  In this paper, we choose  the real-space top-hat window function, in the $k$-space it is 
\begin{equation}\label{window:fun}
	W(k,R_H)=3\left[\frac{\sin\left(kR_H\right)-\left(kR_H\right)
		\cos\left(kR_H\right)}{\left(kR_H\right)^3}\right],
\end{equation}
with the  smoothed scale    $R_H\sim 1/aH$.
The threshold $\delta_c$ of the PBHs formation is dependent on the window function
and the shape of  density perturbations \cite{Young:2020xmk,Musco:2018rwt,Germani:2018jgr}.
For the real space top-hat window function, in this paper, we choose  $\delta_c=0.51$ \cite{Young:2019osy,Musco:2018rwt}.
During radiation domination with constant degrees of freedom, the transfer function  in equation \eqref{variance1} is
\begin{equation}\label{transfer}
	T(k,R_H)=3\left[\frac{\sin\left(\frac{kR_H}{\sqrt{3}}\right)-\left(\frac{kR_H}{\sqrt{3}}\right)
		\cos\left(\frac{kR_H}{\sqrt{3}}\right)}{\left({kR_H}/{\sqrt{3}}\right)^3}\right].
\end{equation}

The masses of primordial black holes in equation \eqref{rho:pbh} obey the   critical scaling law with the formula \cite{Choptuik:1992jv,Evans:1994pj,Niemeyer:1997mt} 
\begin{equation}\label{pbh:mass}
	M_{\text{PBH}}=\kappa M_H(\delta-\delta_c)^{\gamma},
\end{equation}
where $\kappa=3.3$ for the real space top-hat window function and $\gamma=0.36$
in the radiation domination \cite{Choptuik:1992jv,Evans:1994pj}.
The horizon mass related to the horizon scale is
\begin{equation}\label{mass:h}
	M_H\approx 13\left(\frac{g_*}{106.75}\right)^{-1/6}\left(\frac{k}{10^6 \text{Mpc}^{-1}}\right)^{-2}M_\odot,
\end{equation}
where $g_*$ is the number of relativistic degrees of freedom at the formation.
With the help of the background equations of the energy density during radiation domination,  $\rho_b\propto a^{-4}$ and  $\rho_\text{PBH}\propto a^{-3}$,
we obtain the relation of the density parameter of the PBHs  at present  and fraction of PBHs in the Universe at formation  \cite{Byrnes:2018clq},
\begin{equation}\label{beta:omega}
	\Omega_\text{PBH}=\int_{M_\text{min}}^{M_\text{max}} d \ln M_H \left(\frac{M_{eq}}{M_H}\right)^{1/2}\beta(M_H),
\end{equation}
where  $M_{eq}=2.8\times 10^{17}M_{\odot}$ is the horizon mass at the matter-radiation equality.  In our model,  $\beta(M_H)\rightarrow 0$ at the condition $M_H\rightarrow 0$ or $M_H\rightarrow\infty$, so we take the lower limit of integral  as  $M_\text{min}=0$ and the upper limit of that as  $M_\text{max}=\infty$, for the sake of simplicity.
The fraction of primordial black holes in the dark matter at present  is
\begin{equation}\label{fpbh:tot}
	f_{\text{PBH}}=\frac{\Omega_{\text{PBH}}}{\Omega_\text{DM}}=\int f(M_\text{PBH}) d\ln M_\text{PBH},
\end{equation}
where the PBHs mass function is defined as 
\begin{equation}\label{mass:func}
	f(M_\text{PBH})=\frac{1}{\Omega_{\text{DM}}}\frac{d \Omega_{\text{PBH}}}{d \ln M_{\text{PBH}}}.
\end{equation}
Combining relation  \eqref{beta:omega} and definition  \eqref{mass:func},  using equation  \eqref{pbh:mass} and $d\delta/d\ln M_{\text{PBH}}=\mu^{1/\gamma}/\gamma$, the mass function \eqref{mass:func}  becomes \cite{Byrnes:2018clq}
\begin{equation}\label{app:fpbh:beta}
	\begin{split}
		f(M_\text{PBH})&=\frac{1}{\Omega_{\text{DM}}} \int_{M_\text{min}}^{M_\text{max}}\frac{d M_H}{M_H}
		\frac{M_\text{PBH}}{\gamma M_H} \sqrt{\frac{M_{eq}}{M_H}}\\
		&\times\frac{1}{3\pi} \left(\frac{\sigma_1}{\sqrt{3}\sigma_0 aH}\right)^3\frac{1}{\sigma_0^4}
		\left(\mu^{1/\gamma}+\delta_c\right)^3 \times \mu^{1/\gamma} \exp\left[-\frac{\left(\mu^{1/\gamma}+\delta_c\right)^2}{2\sigma_0^2}\right],
	\end{split}
\end{equation}
with $\mu=M_{\text{PBH}}/(\kappa M_H)$.

\begin{figure}[htbp]
	\centering
	\includegraphics[width=0.95\columnwidth]{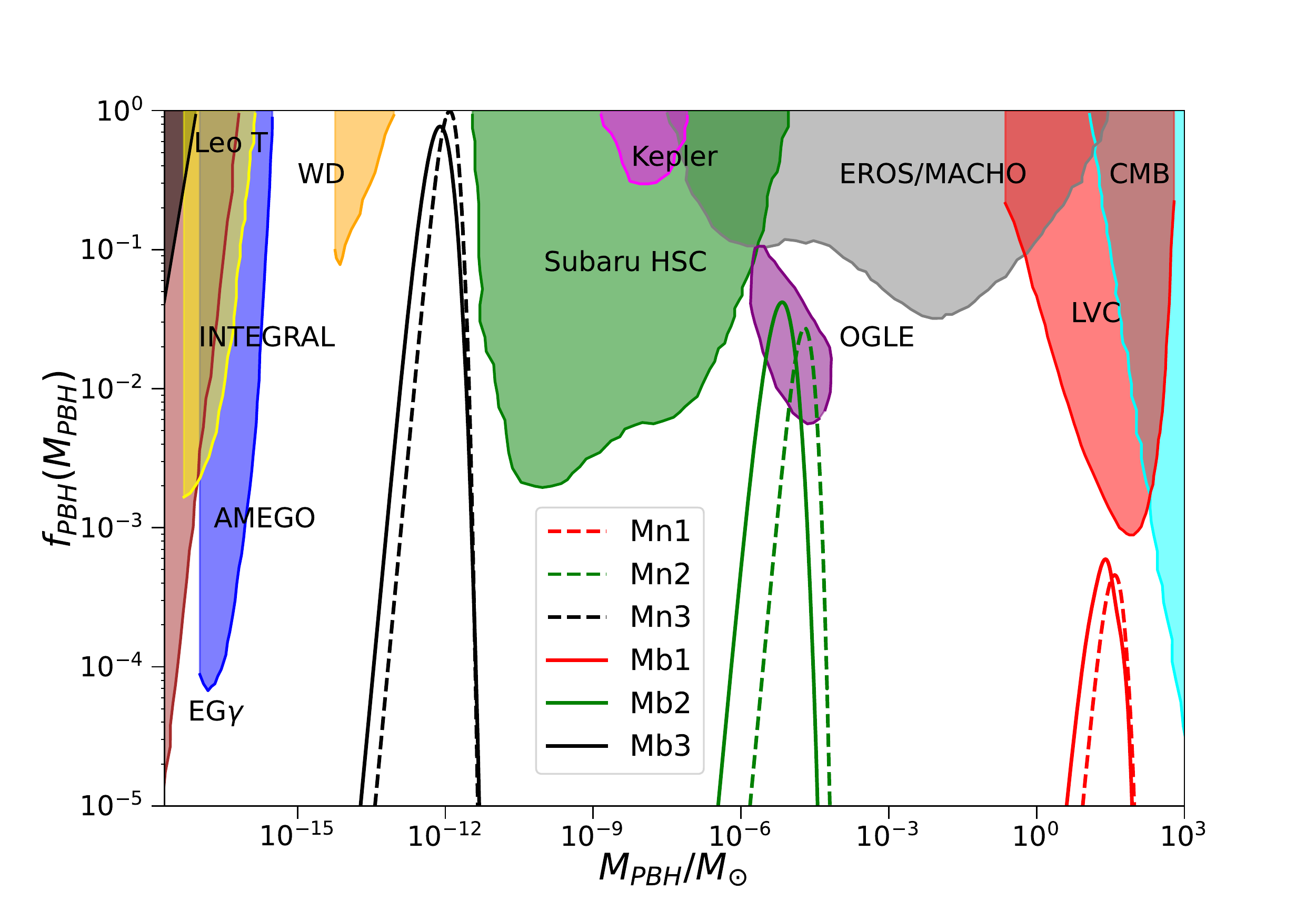}
	\caption{The curves represent PBH  mass functions of the models listed in table \ref{tab:pr}. 
The most comprehensive constraints on non-evaporated PBHs abundance are displayed in  figure 10 of Ref. \cite{Carr:2020gox}, with the mass of PBHs covering from about $10^{-19}M_\odot$ to $10^{22} M_\odot$.  
In this paper, we  are interested in the constraints on the PBHs with masses from about $10^{-17}M_\odot$ to $10^3 M_\odot$;   
and the  shaded regions show the main observational constraints on the PBH abundance:
the cyan region from accretion constraints by CMB \cite{Ali-Haimoud:2016mbv,Poulin:2017bwe},
the red region from LIGO-Virgo Collaboration measurements \cite{Ali-Haimoud:2017rtz,Raidal:2018bbj,Vaskonen:2019jpv,DeLuca:2020qqa,Wong:2020yig,Hutsi:2020sol},
		the gray region from the EROS/MACHO \cite{Tisserand:2006zx},
		the green region from microlensing events with Subaru HSC \cite{Niikura:2017zjd},
		the magenta region from the Kepler satellite \cite{Griest:2013esa}, the orange region from white dwarf explosion (WD) \cite{Graham:2015apa}, the blue region from near future MeV telescope AMEGO \cite{Ray:2021mxu}, the yellow region from galactic center 511 keV gamma-ray line (INTEGRAL) \cite{Laha:2019ssq,Dasgupta:2019cae,Laha:2020ivk}, the black region from Leo T gas heating \cite{Laha:2020vhg}, 
		the brown region from extragalactic gamma-rays by PBH evaporation (EG$\gamma$) \cite{Carr:2009jm}.  The purple region are the allowed PBH abundance from the ultrashort-timescale microlensing events in the OGLE data \cite{Niikura:2019kqi}. }\label{fig:fpbh}
\end{figure}

Using the numerical results of the power spectra of the models listed in table \ref{tab:pr},  combining equations \eqref{variance1} and \eqref{app:fpbh:beta},
we obtain the mass function of the PBHs  and the results are displayed in figure \ref{fig:fpbh},  the corresponding   PBHs abundance $f_\text{PBH}$ and PBHs masses  $M_\text{peak}$ at the peak are listed in table \ref{tab:pbh}.  The PBHs with stellar masses, planetary masses, and $10^{-12} M_\odot$ are produced and denoted by red lines, green lines, and black lines in figure \ref{fig:fpbh}, respectively.  The PBHs with stellar masses, labeled as  $``1"$ in  table \ref{tab:pbh},  can explain the black holes in LIGO/Virgo events \cite{Bird:2016dcv,Sasaki:2016jop, Abbott:2016blz}. The PBHs with planetary masses, labeled as  ``$2$" in  table \ref{tab:pbh}, can explain the ultrashort-timescale microlensing events in   OGLE data  \cite{Niikura:2019kqi}  and the anomalous orbits of trans-Neptunian objects\cite{Scholtz:2019csj}.
The PBHs with masses around  $10^{-12} M_\odot$, labeled as  ``$3$" in  table \ref{tab:pbh}, can account for   almost all of the dark matter,  and their abundances 
are  $f_\text{PBH}=0.95$ for model   ``Mn3"   and  $f_\text{PBH}=0.96$ for model ``Mb3".

\begin{table}[htbp]
	\renewcommand\tabcolsep{5.0pt}
	\centering
	\begin{tabular}{lllllllllll}
		\hline
		\hline
		Model \quad   &$\mathcal{P}_{\zeta(\text{peak})}$& $M_\text{peak}/M_\odot$&$f_\text{PBH}$& $f_c/\text{Hz}$\\
		\hline
		Mn1 \quad   &$5.18\times 10^{-3}$&$38$&$4.71\times 10^{-4}$&$ 1.83\times 10^{-9}$\\
		Mn2 \quad   &$4.75\times 10^{-3}$&$1.95\times 10^{-5}$&$2.7\times 10^{-2}$&$ 2.52\times 10^{-6}$\\
		Mn3 \quad   &$4.14\times 10^{-3}$&$1.23\times10^{-12}$& $0.95$ & $9.76\times10^{-3}$\\
		\hline
		Mb1 \quad   &$4.59\times 10^{-3} $&$25$&$7.12\times10^{-4}$&$1.95\times 10^{-9}$\\
		Mb2 \quad   &$4.11\times 10^{-3}$&$6.93\times 10^{-6}$&$5.39\times 10^{-2}$&$ 4.01\times 10^{-6}$\\
		Mb3 \quad   &$3.54\times 10^{-3}$&$7.87\times 10^{-13}$&$0.96 $&$1.07\times10^{-2}$\\
		\hline
		\hline
	\end{tabular}
	\caption{The results for the peak amplitude of   primordial scalar power spectrum, the   abundance and peak mass  of PBHs, and the peak frequency of SIGWs for the inflation models with the chosen parameter sets listed in table \ref{tab:pr}.}
	\label{tab:pbh}
\end{table}

\subsection{scalar-induced gravitational waves}
In addition to supplying the condition of the formation of PBHs,  during radiation domination and after reentering the horizon, 
the large scalar perturbations  can  induce  the gravitational waves with   frequencies  ranging from nHz to mHz. 
The SIGWs with nHz can be detected by  PTA and account for the NANOGrav 12.5yrs signal, and  those  with mHz
 can be detected by the space-based GW detectors like
LISA, Taiji, and TianQin in the future.
In  the cosmological background and neglecting the anisotropic
stress,  the perturbed metric in  the  Newtonian gauge is
\begin{equation}
	\begin{split}
		d s^2=-a^2(\eta)(1+2\Phi)d\eta^2 +a^2(\eta)\left[(1-2\Phi)\delta_{ij}+\frac12h_{ij}\right]d x^i d x^j,
	\end{split}
\end{equation}
where $\eta$ is the conformal time, $\Phi$ is the Bardeen potential. 
The  tensor perturbations $h_{ij}$ expressed in the   Fourier space are
\begin{equation}
	\label{hijkeq1}
	h_{ij}(\bm{x},\eta)=\int\frac{  d^3k  e^{i\bm{k}\cdot\bm{x}}}{(2\pi)^{3/2}}
	[h_{\bm{k}}(\eta)e_{ij}(\bm{k})+\tilde{h}_{\bm{k}}(\eta)\tilde{e}_{ij}(\bm{k})],
\end{equation}
where $e_{ij}(\bm{k})$ and $\tilde{e}_{ij}(\bm{k})$  are the plus and cross polarization tensors which can be expressed as
\begin{gather}
	e_{ij}(\bm{k})=\frac{1}{\sqrt{2}}\left[e_i(\bm{k})e_j(\bm{k})-\tilde{e}_i(\bm{k})\tilde{e}_j(\bm{k})\right], \\
	\tilde{e}_{ij}(\bm{k})=\frac{1}{\sqrt{2}}\left[e_i(\bm{k})\tilde{e}_j(\bm{k})+\tilde{e}_i(\bm{k})e_j(\bm{k})\right],
\end{gather}
with $\bm e\cdot \tilde{\bm e}=\bm e \cdot \bm{k}= \tilde{\bm e}\cdot\bm{k}$.

For either polarization, the tensor perturbations induced from linear scalar perturbations in the Fourier space  satisfy
\cite{Ananda:2006af,Baumann:2007zm}
\begin{equation}
	\label{eq:hk}
	h''_{\bm{k}}+2\mathcal{H}h'_{\bm{k}}+k^2h_{\bm{k}}=4S_{\bm{k}},
\end{equation}
 where a prime denotes the derivative with respect to the conformal time,
$h'_{\bm{k}}=dh_{\bm{k}}/d\eta$, and $\mathcal{H}=a'/a $ is the conformal Hubble parameter, $S_{\bm{k}}$ is the second order   source from the linear scalar perturbations,
\begin{equation}
	\label{hksource}
	\begin{split}
		S_{\bm{k}}=\int \frac{d^3\tilde{k}}{(2\pi)^{3/2}}e_{ij}(\bm{k})\tilde{k}^i\tilde{k}^j
		\left[2\Phi_{\tilde{\bm{k}}}\Phi_{\bm{k}-\tilde{\bm{k}}} + \frac{1}{\mathcal{H}^2} \left(\Phi'_{\tilde{\bm{k}}}+\mathcal{H}\Phi_{\tilde{\bm{k}}}\right)
		\left(\Phi'_{\bm{k}-\tilde{\bm{k}}}+\mathcal{H}\Phi_{\bm{k}-\tilde{\bm{k}}}\right)\right].
	\end{split}
\end{equation}
The relation between  Bardeen potential  $\Phi_{\bm{k}}$ and  the primordial curvature perturbation $\zeta_{\bm{k}}$  in  Fourier space is
\begin{equation}
	\Phi_{\bm{k}}=\frac{3+3w}{5+3w}T(k,\eta) \zeta_{\bm{k}},
\end{equation}
where $T(k,\eta)$ is the transfer function \eqref{transfer}.
The definition of the power spectrum $\mathcal{P}_h(k,\eta)$ for the SIGWs is
\begin{equation}
	\label{eq:pwrh}
	\langle h_{\bm{k}}(\eta)h_{\tilde{\bm{k}}}(\eta)\rangle
	=\frac{2\pi^2}{k^3}\delta^{(3)}(\bm{k}+\tilde{\bm{k}})\mathcal{P}_h(k,\eta).
\end{equation}
The tensor perturbation \eqref{eq:hk} can be solved by the Green function method and the solution is 
\begin{equation}\label{hk:green}
	h_k(\eta)=\frac{4}{a(\eta)}\int_{\eta_k}^{\eta}d \tilde{\eta}g_k(\eta,\tilde{\eta})a(\tilde{\eta})S_k(\tilde{\eta}),
\end{equation}
where the   corresponding  Green function is
\begin{equation}\label{green}
	g_k(\eta,\eta')=\frac{\sin\left[k(\eta-\eta')\right]}{k}.
\end{equation}
Substituting  the result \eqref{hk:green} into  definition \eqref{eq:pwrh}, we obtain  \cite{Baumann:2007zm,Ananda:2006af,Kohri:2018awv,Espinosa:2018eve,Lu:2019sti}
\begin{equation}\label{ph}
	\begin{split}
		\mathcal{P}_h(k,\eta)=
		4\int_{0}^{\infty}dv\int_{|1-v|}^{1+v}du \left[\frac{4v^2-(1-u^2+v^2)^2}{4uv}\right]^2   I_{RD}^2(u,v,x)\mathcal{P}_{\zeta}(kv)\mathcal{P}_{\zeta}(ku),
	\end{split}
\end{equation}
where $u=|\bm{k}-\tilde{\bm{k}}|/k$, $v=\tilde{k}/k$, $x=k\eta$  and the integral kernel $I_{\text{RD}}$  is
\begin{equation}
	\label{irdeq1}
	\begin{split}
		I_{\text{RD}}(u, v, x)=&\int_1^x dy\, y \sin(x-y)\{3T(uy)T(vy)\\
		&+y[T(vy)u T'(uy)+v T'(vy) T(uy)]\\
		&+y^2 u v T'(uy) T'(vy)\}.
	\end{split}
\end{equation}
Substituting equation \eqref{ph} into the definition of energy density of   SIGWs,
\begin{equation}
	\label{density}
	\Omega_{\mathrm{GW}}(k,\eta)=\frac{1}{24}\left(\frac{k}{aH}\right)^2\overline{\mathcal{P}_h(k,\eta)},
\end{equation}
we get \cite{Espinosa:2018eve,Lu:2019sti}
\begin{equation}
	\label{SIGWs:gwres1}
	\begin{split}
		\Omega_{\mathrm{GW}}(k,\eta)=&\frac{1}{6}\left(\frac{k}{aH}\right)^2\int_{0}^{\infty}dv\int_{|1-v|}^{1+v}du \left[\frac{4v^2-(1-u^2+v^2)^2}{4uv}\right]^2\\
		&\times\overline{I_{\text{RD}}^{2}(u, v, x)} \mathcal{P}_{\zeta}(kv)\mathcal{P}_{\zeta}(ku),
	\end{split}
\end{equation}
where $\overline{I_{\text{RD}}^{2}}$ is the oscillation time average of the integral kernel.
After  formation during the radiation domination,  the SIGWs  behave  like radiation,
 so  the energy density of the SIGWs is in direct proportion to the energy density of the radiation. Using this property, we can obtain  the energy density of the SIGWs  at present easily and it is
\begin{equation}\label{d}
	\Omega_{\mathrm{GW}}(k,\eta_0)=c_g\Omega_{r,0}  \Omega_{\mathrm{GW}}(k,\eta),
\end{equation}
where $\Omega_{r,0}$ is the energy density of radiation at present, and \cite{Vaskonen:2020lbd,DeLuca:2020agl}
\begin{equation}\label{gwcg}
	c_g=0.387\left(\frac{g_{*,s}^4g_*^{-3}}{106.75}\right)^{-1/3}.
\end{equation}

\begin{figure}[htbp]
	\centering
	\includegraphics[width=0.95\columnwidth]{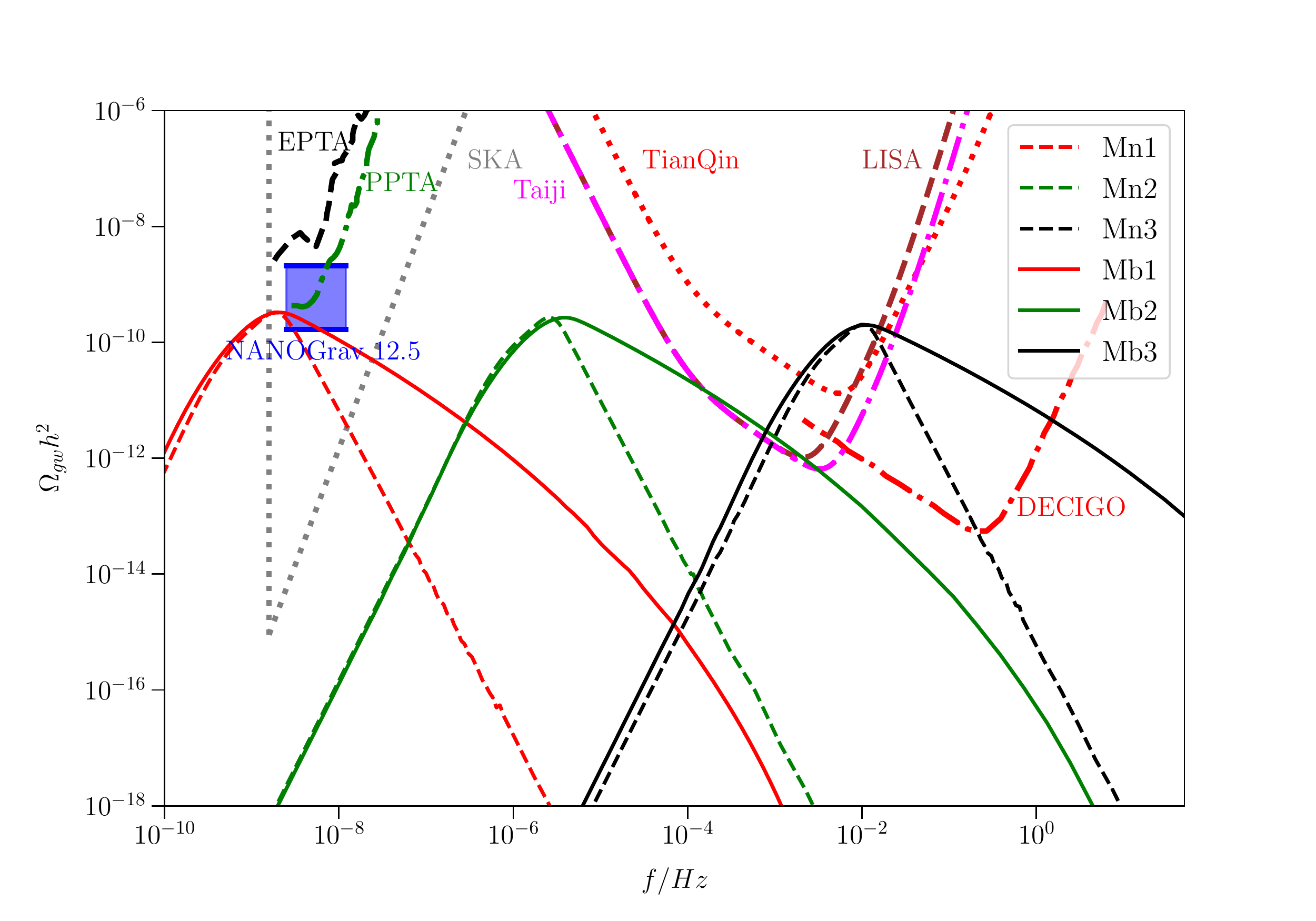}
	\caption{The corresponding  scalar-induced secondary gravitational waves from the models listed in table \ref{tab:pr}.
		The black dashed curve denotes the EPTA limit \cite{Ferdman:2010xq,Hobbs:2009yy,McLaughlin:2013ira,Hobbs:2013aka,Lentati:2015qwp},
		the green dot-dashed curve denotes the PPTA limit \cite{Shannon:2015ect},
		the gray dotted curve denotes the SKA limit \cite{Moore:2014lga},
		the red  dotted curve in the middle denotes the TianQin limit \cite{Luo:2015ght},
		the magenta dot-dashed curve shows the Taiji limit \cite{Hu:2017mde},
		the brown dashed curve shows the LISA limit \cite{Audley:2017drz}, and the red dot-dashed curve in the right shows the DECIGO limit \cite{Kawamura:2011zz}.  }\label{fig:gw}
\end{figure}

Substituting the  numerical results of the power spectra of the models listed in table \ref{tab:pr} into equation \eqref{SIGWs:gwres1},  we obtain the  energy density of  SIGWs and the numerical results are displayed in  figure \ref{fig:gw}, the corresponding  peak frequencies  $f_c$ are listed in table \ref{tab:pbh}.
The SIGWs of models  labeled as ``1" in table  \ref{tab:pbh} are denoted by the red lines in figure \ref{fig:gw} with   frequencies around $10^{-9}$ Hz. They are consistent with 
 $2\sigma $ region of the NANOGrav 12.5 yrs signal, which indicates that the NANOGrav 12.5 yrs signal may be  SIGWs.
The SIGWs of models  labeled as ``2" in table  \ref{tab:pbh} are denoted by the green lines in figure  \ref{fig:gw} with frequencies around $10^{-6}$ Hz. The broad peak  can be detected by the space-based detectors LISA and Taiji.   
The SIGWs of models  labeled as ``3" in table \ref{tab:pbh} are denoted by the black lines in figure   \ref{fig:gw} with  frequencies around $10^{-3}$ Hz, and  can be   detected by the   LISA, Taiji, TianQin, and DECIGO in the future. 

\section{Conclusion}
PBHs and SIGWs can be produced from the inflation models with a transient ultra-slow-roll process, where the equation of motion for the scalar field is  $\ddot{\phi}+3 H\dot{\phi}\approx 0$.  
For the scalar-tensor inflation, the ultra-slow-roll condition can be realized by 
taking $\Omega^2/[\omega+3\Omega'^2/(2\Omega)]\ll1$ with $\Omega(\phi)$  coupling to the Ricci scalar and $\omega(\phi)$   to the kinetic term.
For the coupling function $\Omega(\phi)$ with quadratic form  $\Omega(\phi)=1-2 \phi/\phi_c  +(1+\delta^2) \left({\phi}/{\phi_c}\right)^2$, 
 under the condition  $\delta^2\ll1$, the ultra-slow condition is satisfied  at the point $\phi_c$  and the power spectra can be enhanced enough to 
produce PBHs and generate SIGWs. 
The parameter $\phi_c$ controls the position of the peak in  power spectra and also governs the masses of   PBHs and the frequencies of  SIGWs. The kinetic coupling function is $\omega(\phi)+{3 \Omega'^2}/{(2\Omega)}=a \Omega^{b} + c \Omega$, determining the  shape of the peak  in power spectra.

In this paper, we produce three kinds of PBHs:  the PBHs with stellar masses,  those with planetary  masses, and those  with masses around $10^{-12}M_\odot$.  The   cases  with a narrow peak and a broad peak are both given  for each kind.
The first kind PBHs have the peak masses $M_\text{peak}=38 M_{\odot}$ (narrow peak) and  $M_\text{peak}=25 M_{\odot}$ (broad peak), and  may be the  sources of the GWs in the LIGO/Virgo events.  The  corresponding SIGWs have the peak frequencies  around $10^{-9}$ Hz and can explain the NANOGrav 12.5yrs signal. 
The  second  kind PBHs have the peak masses  $M_\text{peak}=1.95\times 10^{-5} M_{\odot}$ (narrow peak) and  $M_\text{peak}=6.93\times 10^{-6} M_{\odot}$ (broad peak),   can  explain  the ultrashort-timescale microlensing events in the OGLE data.  The  corresponding SIGWs have the peak frequencies  around $10^{-6}$ Hz, and the broad peak case can be  detected by  the space-based detectors LISA and Taiji.
The third   kind PBHs have peak masses $M_\text{peak}=1.23\times 10^{-12} M_{\odot}$ (narrow peak) with the PBHs abundance $f_\text{PBH}=0.95$, and  $M_\text{peak}=7.87\times 10^{-13} M_{\odot}$ (broad peak)  with the PBHs abundance $f_\text{PBH}=0.96$, they can account for almost all of the dark matter. 
The  corresponding SIGWs have the  peak frequencies around  $10^{-3}$ Hz and can be detected by  the space-based detectors LISA, Taiji,  TianQin, and DECIGO.
The scalar tilt and tensor-to-scalar ratio of these models are about	
$n_s\approx 0.965, \quad r\approx 0.004$ with the $e$-folds $N\approx 60$,  
which are consistent with the Planck 2018 observational data.

In conclusion, the scalar-tensor inflation with the  quadratic form coupling function $\Omega(\phi)=1-2 \phi/\phi_c  +(1+\delta^2) \left({\phi}/{\phi_c}\right)^2$ can successfully enhance the power spectra, produce the  PBHs, and generate the  SIGWs. 
The masses of the PBHs and the frequencies of the SIGWs can be adjusted by the parameter $\phi_c$, and the shape of the peak can  be  adjusted by the coupling function $\omega(\phi)$.

\acknowledgments
We thank Yizhou Lu, Xing-Jiang Zhu, Zu-Cheng Chen, Xiao-Jin Liu,  Zhi-Qiang You, and  Lang Liu for useful discussions.
This research is supported  by the National Natural Science Foundation of China under Grant No. 12205015 and the supporting fund for young researcher of Beijing Normal University under Grant No. 28719/310432102.



\providecommand{\href}[2]{#2}\begingroup\raggedright\endgroup

\end{document}